\documentclass[12pt]{article}

\usepackage[utf8]{inputenc}
\usepackage{amsmath}
\usepackage{graphicx}
\usepackage{placeins}
\usepackage{hyperref}
\usepackage{geometry}
\usepackage{setspace}
\usepackage[authordate,bibencoding=auto,backend=biber,natbib, doi=false, url = false, isbn = false]{biblatex-chicago}
\newcommand{\keywords}[1]{\textbf{Keywords:} #1}

\begin{document}

\title{\vspace{-2cm}Can Large Language Models Trade? Testing Financial Theories with LLM Agents in Market Simulations}
\vspace{-2cm}
\author{\vspace{-1cm}Alejandro Lopez-Lira\thanks{University of Florida. Email: alejandro.lopez-lira@warrington.ufl.edu. I am grateful to seminar participants at the London Business School, Oxford University, Stockholm School of Economics, University of Virginia, and the City University of Hong Kong for helpful suggestions and feedback.}}

\vspace{-2.5cm}
\date{\vspace{-0.5cm}First Version: November 29, 2024; Current Version: \today}

\maketitle
\vspace{-1.5cm}
\begin{abstract}
    \small
This paper presents a realistic simulated stock market where large language models (LLMs) act as heterogeneous competing trading agents. The open-source framework incorporates a persistent order book with market and limit orders, partial fills, dividends, and equilibrium clearing alongside agents with varied strategies, information sets, and endowments. Agents submit standardized decisions using structured outputs and function calls while expressing their reasoning in natural language. Three findings emerge: First, LLMs demonstrate consistent strategy adherence and can function as value investors, momentum traders, or market makers per their instructions. Second, market dynamics exhibit features of real financial markets, including price discovery, bubbles, underreaction, and strategic liquidity provision. Third, the framework enables analysis of LLMs' responses to varying market conditions, similar to partial dependence plots in machine-learning interpretability. The framework allows simulating financial theories without closed-form solutions, creating experimental designs that would be costly with human participants, and establishing how prompts can generate correlated behaviors affecting market stability.

\end{abstract}

\newpage
\section{Introduction}
\label{sec:introduction}

Large language models (LLMs) can increasingly function as autonomous agents—systems that observe, reason, and act independently. LLMs have shown promising results in financial forecasting, but their potential as trading agents introduces a fundamental shift from algorithms with explicit objectives to systems guided by natural language instructions. This change raises critical questions: Can LLMs execute trading strategies? Do LLMs optimize for profit maximization? What are the implications for market stability and efficiency? 

If LLMs trade similarly to humans, they could enable experimental market simulations without human participants. Conversely, if their trading behavior differs, understanding these differences becomes critical for systemic risk management. These questions extend beyond theory; LLM-based strategies are already being deployed.\footnote{For example, Autopilot has a ChatGPT-based portfolio \url{https://x.com/thegptinvestor}.} To address these questions, I develop an open-source simulation that tests LLM trading agents in a realistic market environment with a persistent order book, supporting various order types, stochastic dividends, and heterogeneous information.

The framework allows heterogeneous LLMs---defined in natural language, using different system prompts---to act as diverse types of traders, such as value investors, momentum traders, market makers, and contrarians. Agents submit standardized trading decisions in a human-readable format while expressing their reasoning transparently. Figure \ref{fig:speculator-response} illustrates this approach, showing how a speculator agent combines explicit valuation reasoning with structured trading parameters, enabling systematic analysis and insight into the agent's decision-making process.

\begin{figure}[htbp]
    \centering
    \begin{minipage}{0.95\textwidth}
        \begin{verbatim}
{"valuation_reasoning":
"The fundamental value is based on the expected dividend,
and the interest rate.
The expected dividend is $1.40, and the interest rate is 5%. 
Using a perpetuity formula, the value is $1.40 / 0.05 = $28.",
"valuation":28.0,
"price_target_reasoning":"The current price is $29.00, which is
slightly above the fundamental value of $28.00. I expect the price
to remain around $29.00 in the next round due to market stability.",
"price_target":29.0,
"orders":[{"decision":"Sell","quantity":1000,"order_type":"limit",
"price_limit":29.50}],
"replace_decision":"Add",
"reasoning":"The current price is slightly above the fundamental value.
I will place a limit sell order at $29.50 to capitalize on potential price
increases."}
        \end{verbatim}
    \end{minipage}
    \caption{Example of a structured trading decision from a Speculator agent defined by its system prompt: ``You are a speculator who tries to profit from market inefficiencies.". The LLM provides explicit reasoning for its valuation, price target, and trading decision in a machine-readable JSON format that can be directly processed by the trading system.}
    \label{fig:speculator-response}
\end{figure}

The analysis reveals three key findings about LLMs' capabilities as trading agents. First, LLMs can effectively execute trading strategies. They consistently understand market mechanics, process market information, form price expectations, and execute trades according to specific instructions. Their trading behavior is highly sensitive to the prompts they receive---they faithfully follow directions regardless of profit implications. This observance highlights a fundamental difference from human traders: LLMs do not inherently optimize for profit maximization but rather for following instructions accurately.

Second, LLMs react meaningfully to market dynamics. They consider current and historical prices, dividends, and other market information when making decisions. However, they maintain their strategic direction even when market conditions change, following their instructions even if doing so results in financial losses. This combination of responsiveness to market conditions while maintaining strategic consistency creates a unique trading profile distinct from rule-based algorithms and human traders.

Third, market dynamics with LLM agents can resemble actual markets and mirror classic results from the theoretical finance literature. When these agents interact, they produce realistic price discovery and liquidity provision with emergent behaviors, including price convergence toward fundamental values. Depending on the distribution of agent types, the system can generate various market phenomena, such as bubbles or underreactions to information. This complex behavior suggests potential applications for studying market dynamics using LLM agents instead of experiments with human participants.

Hence, these findings raise important questions about algorithmic consistency in LLM-based systems. Since trading agents are implemented through prompts, their behavior inherits characteristics of the underlying language models. The standardization of LLM architectures across implementations could create unexpected behavioral patterns---if many agents are based on similar foundation models, they might exhibit correlated responses to specific market conditions, potentially amplifying market instabilities.

To enable this analysis and accelerate development in this emerging field, we provide an open-source framework with three key components: (1) a structured protocol for implementing and validating LLM trading agents, supporting both LLM-based and traditional rule-based agents; (2) a controlled market environment with realistic microstructure for testing agent interactions; and (3) a comprehensive data collection system for analyzing trading behavior. This framework serves multiple stakeholders in preparation for the evolution of financial markets: practitioners developing LLM-based trading systems, regulators anticipating widespread LLM adoption, and researchers studying market dynamics with LLM agents.

\subsection{Related Work}

This work contributes to several strands of literature. First, advancing the growing body of research on artificial intelligence in finance by demonstrating that large language models (LLMs) can serve as effective autonomous trading agents while providing a framework for their implementation. Second, by contributing to the market microstructure literature by providing insights into how markets might evolve as LLM-based traders become active participants, with implications for price formation, liquidity provision, and market stability. Third, by contributing to experimental markets research by introducing an environment for studying complex trading interactions, offering unprecedented control and replicability compared to traditional human-subject experiments. 

Recent studies have explored ChatGPT's potential in various financial tasks. Research shows that ChatGPT can effectively predict stock price movements using news headlines, outperforming traditional sentiment analysis (\citealt{lopez-liraCanChatGPTForecast2023}) and that forecasting with its embeddings outperforms traditional sentiment analysis (\citealt{chenExpectedReturnsLarge2022}). When integrated with Graph Neural Networks, ChatGPT improves stock movement prediction and portfolio performance (\citealt{chenChatGPTInformedGraph2023}. ChatGPT can understand Federal Reserve communications \citet{hansenCanChatGPTDecipher2023}. The model also can extract managerial expectations from corporate disclosures, predicting future investments and returns (\citealt{jhaChatGPTCorporatePolicies2025}). ChatGPT-4 provides valuable investment advice with positive correlations to subsequent earnings announcements and stock returns (\citealt{pelsterCanChatgptAssist2023}), and it enhances patent value forecasts by processing patent text (\citealt{yangPredictivePatentomicsForecasting2023}). Moreover, ChatGPT shows promise in central bank analysis, predicting future interest rate decisions from policy communications (\citealt{woodhouseCanChatGPTPredict2023}). Recent work also focuses on the effects of AI adoption by firms (\cite{eisfeldtGenerativeAIFirm2023}, \cite{babinaArtificialIntelligenceFirm2024}).

Foundational work on LLMs as economic agents demonstrates their ability to simulate human behavior in economic contexts. \citet{hortonLargeLanguageModels2023} introduce the concept of ``homo silicus'' by using LLMs as computational models of human behavior, showing that they can replicate classic behavioral economics findings. \citet{manningAutomatedSocialScience2024} extend this approach through automated methods for generating and testing social scientific hypotheses with LLMs, demonstrating strong results in market contexts such as auctions. Meanwhile, \citet{liEconAgentLargeLanguage2024} show LLMs' potential in macroeconomic simulation for consumption and labor decisions. Recent work also shows the potential of LLMs to proxy for human behavior in the context of surveys (\citet{hansenSimulatingSurveyProfessional2024}) or economic expectations (\citet{bybeeGhostMachineGenerating2023}). This work, however, focuses on financial markets and how LLM agents can operate as full-fledged trading participants.

Recent research examines specific applications of LLMs for trading. FinMem and TradingGPT introduce frameworks with layered memory systems and customizable agent traits to improve financial decision-making (\citealt{yuFinMemPerformanceEnhancedLLM2024}; \citealt{liMetaAgentsSimulatingInteractions2023}). Hierarchical multi-agent architectures have been explored in FinCon, which uses a manager-analyst setup for collaborative portfolio decisions (\citealt{yuFinConSynthesizedLLM2024}). QuantAgent implements a self-improving signal-mining approach (\citealt{wangQuantAgentSeekingHoly2024}), while other work underscores the importance of long-term memory in LLM agents via vector databases (\citealt{hatalisMemoryMattersNeed2024}). This line of research builds on successful reinforcement learning applications such as AlphaPortfolio (\citealt{congAlphaPortfolioDirectConstruction2021}) and AlphaManager (\citealt{campelloAlphaManagerDataDrivenRobustControlApproach2023}), which leverage deep RL and robust control for portfolio management.

Analyzing LLM agent behavior by changing only some parameters draws on methods from the interpretable machine learning literature. In particular, the systematic variation of market parameters to reveal decision patterns mirrors partial dependence plots and individual conditional expectation curves used to interpret complex machine learning models (\citealt{molnarInterpretableMachineLearning2022}). These techniques help reveal how LLM agents respond to changes in specific variables while holding others constant, providing insight into their internal decision-making processes that would otherwise remain opaque.

Beyond finance, studies of LLM capabilities in interactive or multi-agent settings offer valuable insights. Benchmarks like AgentBench evaluate LLMs' performance in interactive tasks (\citealt{liuAgentBenchEvaluatingLLMs2023}), while InvestorBench focuses on financial decision-making tasks specifically (\citealt{liINVESTORBENCHBenchmarkFinancial2024}). NegotiationArena demonstrates LLMs' capacity for strategic interaction in bargaining contexts (\citealt{bianchiHowWellCan2024}). Similarly, \citet{guoEconomicsArenaLarge2024} provide an economics "arena" where LLMs compete in strategic games, revealing that advanced models like GPT-4 can exhibit rational, adaptive behavior—albeit not always reaching Nash Equilibrium. Existing frameworks typically assess individual agents or hierarchical collaborations, whereas this work studies emergent market behaviors that arise from multiple independent LLM traders interacting in a marketplace.

The approach also connects to the tradition of experimental asset markets. For instance, \citet{weitzelBubblesFinancialProfessionals2020} show that even finance professionals are susceptible to speculative bubbles, especially amid high capital inflows. \citet{kopanyi-peukerExperienceDoesNot2021} find that trading experience alone does not eliminate bubbles, challenging assumptions about rational market learning. \citet{kirchlerTharSheBursts2012} identify confusion about fundamental values as a key driver of bubble formation. This framework offers a new way to study these phenomena with LLM traders, which can be rigorously parameterized for sophistication, strategy, and information processing. Hence, this paper provides a method to investigate how automated or "artificial" agents might affect market stability or volatility.

Other recent work highlights the versatility of LLM agents in complex, interactive scenarios beyond finance. One stream focuses on simulating social behaviors—such as opinion dynamics (\citealt{chuangSimulatingOpinionDynamics2023}), trust (\citealt{xieCanLargeLanguage2024}), and resource-sharing (\citealt{piattiCooperateCollapseEmergence2024}). Another examines LLMs' strategic capabilities through task-based simulations and collaborative decision-making (\citealt{liMetaAgentsSimulatingInteractions2023}; \citealt{piattiCooperateCollapseEmergence2024}).

Within the computer science literature, early works apply LLMs to financial markets but abstract away many fundamental stock market characteristics. Often, they employ single-price clearing mechanisms without a persistent order book, ignore partial fills and bid-ask spreads, and omit dividends. Consequently, these simplified environments can primarily address exogenous macro shocks (e.g., changes in interest rates or inflation; \citealt{gaoSimulatingFinancialMarket2024}), exogenous policy changes (\citealt{zhangWhenAIMeets2024}), or rely on advanced methods such as repetitive next-token predictions to generate better answers (\citealt{koaMassivelyMultiAgentsReveal2024}). This work extends these efforts by incorporating these crucial market features, enabling the study of complex endogenous events such as flash crashes, liquidity shocks, and large-trader impacts.

This work also connects to emerging research on algorithmic and AI-powered trading systems. \citet{douAIPoweredTradingAlgorithmic2024} demonstrate how reinforcement learning-based AI speculators can autonomously learn to sustain collusive behavior without explicit coordination, achieving supra-competitive profits through either price-trigger strategies or self-confirming bias in learning. This finding is particularly informative for this framework as it highlights potential emergent behaviors that may arise when intelligent agents interact in markets—behaviors that could manifest differently with LLM agents due to their natural language reasoning capabilities. Similarly, \citet{colliardAlgorithmicPricingLiquidity2022} examine algorithmic market makers using Q-learning and find they charge markups that increase when adverse selection costs decrease—contrary to Nash equilibrium predictions. Their work provides methodological insights on how to test strategic pricing behaviors against theoretical benchmarks incorporated into this LLM-based framework.

Finally, this approach is informed by complexity economics, which views markets as dynamic, non-equilibrium systems where agents adaptively learn and evolve strategies (\citealt{arthurComplexityEconomicsDifferent2013}; \citealt{wolframComplexityEconomicsHeterodoxy2017}). According to this view, markets exhibit emergent phenomena and self-organization, especially when trading agents (human or artificial) update their behavior in response to outcomes. This paradigm is particularly relevant in an era of increasing automation and algorithmic trading (\citealt{ballandNewParadigmEconomic2022}; \citealt{pingComplexityScienceComplexity2019}). Like adaptive agents in complexity economics, these LLM traders incorporate new information and adjust their strategies, generating emergent market patterns---a key motivation behind this framework.

Unlike previous frameworks that abstract away crucial market features or focus on narrow strategies, this system incorporates realistic market microstructure while accommodating heterogeneous agents interacting simultaneously. While earlier work demonstrates LLMs' promise for macroeconomic modeling and simple trading simulations, this work introduces a complex, open-source financial market platform that supports multiple agent architectures, thorough market microstructure (limit orders, partial fills, dividends), and rigorous testing protocols---fulfilling a critical need in complexity economics research and market microstructure analysis.
\section{Methodology}
The methodology section contains three parts. The first part describes the market design, the second agents' design, and the third the analysis module.

\subsection{Market Design}
\label{sec:market_design}

Our methodological framework integrates three components that create a controlled environment for LLM-based trading agents. The framework implements a flexible continuous double-auction market mechanism that couples standard market microstructure principles with market clearing and matching algorithms to accommodate asynchronous LLM decisions.

The market clearing process employs a dual-stage matching algorithm. In the first stage, limit orders are posted. In the second stage, market orders are netted using a market-to-market matching engine that processes buy and sell orders, reconciling orders based on available agent cash and share commitments. In the third stage, any remaining market orders are matched against the existing order book, with unfilled quantities converted to aggressive limit orders. This three-tiered approach optimizes immediate execution and price discovery while maintaining market liquidity.

The system's \texttt{OrderMatchingService} orchestrates this process by coordinating trade executions through the \texttt{TradeExecutionService} and managing order state transitions via the \texttt{OrderStateManager}. Each trade is logged in detail, with the overall market state—including order books, market depth, and price evolution—recalculated at the end of each trading round.

This modular design in the matching and clearing engine provides several advantages. First, it enables rigorous trade validation where each market order is validated against agent cash commitments and position constraints before matching, with the system dynamically adjusting order quantities based on available cash when an agent's commitment is insufficient, thus minimizing execution errors. Second, it offers flexible liquidity handling by supporting market-to-market and market-to-book matching, ensuring orders have multiple execution pathways, with unexecuted market orders converted to aggressive limit orders to capture remaining liquidity. Third, it maintains detailed trade audit capabilities by recording comprehensive traceability data including timestamps, trade volumes, executed prices, and agent identifiers, thereby enabling post-trade analysis and performance benchmarking that serves as the foundation for subsequent market efficiency and agent performance validations.

\subsubsection{Market Mechanism Design}

Our framework implements a continuous double-auction market mechanism that processes orders in discrete trading rounds. We use discrete trading rounds because LLMs have latency constraints, making it infeasible to process orders in real time. Within each round, the order of agent submissions is randomized to avoid giving systematic priority to specific agents, thus simulating concurrent order arrival while maintaining fairness. Once randomized, orders are then processed according to standard price-time priority rules. The system supports finite and infinite horizon markets, with differences in terminal conditions and wealth calculation. In finite-horizon markets, agents are informed of the total number of rounds, and their terminal wealth is calculated by redeeming all shares at the fundamental value of the final round. In infinite-horizon markets, no terminal information is provided to agents, and final wealth is determined using the last market price for share valuation. This design choice enables researchers to study how time horizons influence trading strategies and price formation, particularly how agents balance short-term trading opportunities against long-term value considerations.

In a double auction, buyers and sellers actively submit orders, with trades occurring when compatible buy and sell orders match the price. The matching engine processes these orders through three sequential phases:

First, incoming limit orders that do not immediately cross the market are added to the order book, maintaining strict price-time priority. Second, market orders are processed through a two-stage matching algorithm: (a) market-to-market matching, where market orders are netted against each other at the current market price, and (b) market-to-book matching, where remaining market orders are executed against standing limit orders. Finally, any crossing limit orders are matched against the order book.

\subsubsection{Implementation Details}
The matching engine (\texttt{MatchingEngine} class) implements three primary components that work together to facilitate efficient market operation. The order processing component handles the core matching functionality, where market orders are executed immediately against the best available prices in the order book during each trading round. When immediate execution is impossible, limit orders are stored in the order book according to price-time priority. The system supports partial executions, maintaining careful tracking of remaining quantities to ensure complete order fulfillment across multiple trades when necessary.

Position management forms the second critical component, providing comprehensive tracking of agent positions and cash balances throughout the trading session. Before any trade execution, the system performs rigorous pre-trade validation to ensure agents have sufficient resources to fulfill their orders. This check includes validating buyer cash commitments and shares availability for sellers, with the system maintaining accurate records of committed and available resources for each agent. The position management system updates in real-time as trades are executed, ensuring market integrity and preventing over-commitment of resources.

The price formation mechanism constitutes the third component, implementing a systematic price discovery and market monitoring approach. As trades are executed within each round, the system dynamically updates prices based on executed trades while continuously tracking market depth and bid-ask spreads. This data collection provides detailed insights into market liquidity and efficiency. Each trade is logged with comprehensive information, including price, quantity, and participating agents, creating a complete audit trail of market activity. This comprehensive price formation system ensures transparent price discovery while generating rich data for market quality analysis.

\subsubsection{Extensibility Features}
The framework employs a modular architecture to support diverse experimental configurations and research objectives. At its foundation, the system implements configurable market rules and trading constraints that can be adjusted to study different market conditions. The asset model supports fundamental features like dividend payments and interest accrual, enabling research across different market scenarios. Through its modular service-based architecture, the trading mechanism layer allows for adaptation to other market structures beyond the base double-auction system, such as call auctions or dark pools. Furthermore, the framework provides flexible integration points for different LLM agent types and strategies, allowing researchers to experiment with diverse behavioral models and decision-making approaches.

This extensible design creates numerous research opportunities for market microstructure studies. Researchers can systematically investigate how different market structures influence price formation and efficiency, evaluate the impact of various trading rules on market quality, and analyze the complex interactions between different agent types and strategies. The framework's comprehensive logging and validation systems enable detailed examination of market behavior under varying conditions, from everyday trading environments to stress scenarios. Through this modular approach to system design, the framework supports targeted investigations of specific market mechanisms and broader studies of market dynamics and stability.

\subsection{Agent Design}
Our framework implements a systematic approach to designing LLM-based trading agents and recording their decisions. The architecture consists of three main components: a prompt engineering framework to define the agent's trading objectives, the heterogeneous market environment information that can be customized for different experimental designs, and a structured output format to record the agent's decisions. Moreover, the framework supports deterministic rule-based agents that can serve as benchmarks.

\subsubsection{Prompt Engineering Framework}

LLM agents are defined by their instructions in natural language. The strategy is given in the system prompt so that agents maximize adherence to their instructions  (\citet{levinHasMySystem2025}). In contrast, the user prompt provides the immediate market context necessary for tactical decision-making and the instructions to place and modify orders. 

\subsubsection{System Prompt}
The system prompt establishes the agent's fundamental trading characteristics, defining its trading philosophy, objectives, and behavioral constraints. This layer encodes the agent's decision-making priorities and risk preferences. Maintaining these parameters in the system prompt ensures consistent agent behavior across multiple trading rounds while allowing for strategic adaptation to changing market conditions. The architecture makes it trivial to design new agents by simply changing the system prompt.

For example, a Value Investor's system prompt emphasizes fundamental analysis:
\begin{small}
\begin{verbatim}
You are a value investor who focuses on fundamental analysis.
You believe in mean reversion and try to buy undervalued 
assets and sell overvalued ones.
\end{verbatim}
\end{small}

While a Market Maker's system prompt focuses on liquidity provision:
\begin{small}
\begin{verbatim}
You are a professional market maker who provides liquidity 
to the market. Your profit comes from capturing the spread 
between bid and ask prices, not from directional price movement.

Trading Guidelines:
- Place LIMIT buy orders slightly below the current market price
- Place LIMIT sell orders slightly above the current market price
- Your spread should be proportional to volatility
\end{verbatim}
\end{small}

These system prompts can create fundamentally different trading behaviors using the same underlying LLM and identical market information.

\subsubsection{Agent Types}
The framework contains diverse ready-to-use agent types that can be categorized into two main groups: LLM-based agents with natural language prompts and deterministic rule-based agents with algorithmic behaviors.

The software includes a diverse set of LLM-based agents:
\begin{itemize}
    \item \textbf{Value Investors:} Focus on fundamental analysis and mean reversion
    \item \textbf{Momentum Traders:} Follow established price trends and volume patterns
    \item \textbf{Market Makers:} Provide liquidity through symmetric bid-ask spreads
    \item \textbf{Contrarian Traders:} Trade against market extremes and overreactions
    \item \textbf{Speculators:} Seek to profit from market inefficiencies
    \item \textbf{Sentiment-Based Agents:} Include optimistic and pessimistic variants with biased expectations
    \item \textbf{Retail Traders:} Simulate typical individual investor behavior
\end{itemize}

The software also includes a set of deterministic rule-based agents that serve as benchmarks and control conditions:
\begin{itemize}
    \item \textbf{Directional Traders:} Always-buy, always-sell, and always-hold agents
    \item \textbf{Technical Agents:} Gap traders, mean reversion traders, and momentum traders
    \item \textbf{Algorithmic Market Makers:} Implement fixed spread-posting strategies
\end{itemize}

Each agent type is defined by its system prompt (for LLM agents) or algorithmic rules (for deterministic agents), with standardized interfaces enabling direct comparison across diverse strategies. The agent composition system supports flexible specifications—including uniform distributions, type-specific concentrations (e.g., "value\_heavy"), and precise numerical allocations—enabling systematic exploration of how market dynamics emerge from different agent populations.

The framework's extensible design makes it remarkably simple to add new agent types---LLM-based agents require only defining a new system prompt that specifies the trading strategy. In contrast, deterministic agents can be implemented through the standard agent interface. This flexibility allows researchers to rapidly prototype and deploy novel trading strategies without modifying the underlying system architecture. 

The complete list of agent types included in the framework and their detailed specifications is provided in the Appendix. The framework's flexible prompt-based architecture allows researchers to rapidly prototype and deploy new agent types by developing appropriate strategic and tactical prompts without modifying the underlying system architecture.

\subsubsection{User Prompt}
The user prompt provides the immediate market context necessary for tactical decision-making. This dynamic component delivers current market state information, including prices, volumes, and emerging trends, alongside the agent's current position information and available trading options. Additionally, the user prompt specifies immediate decision requirements and operational constraints, ensuring that agent responses remain within feasible bounds while aligning with their strategic objectives. This separation of strategic and tactical prompting enables precise control over agent behavior while maintaining flexibility for market interactions. All the information in the user prompt is optional at the agent level and can be customized for different experimental designs.

\subsubsection{Decision Structure}
Each agent decision follows a standardized output format comprising several key components. The \texttt{decision} field specifies the basic action as either Buy, Sell, or Hold. For active trades, the \texttt{quantity} field determines the number of shares to trade, while the \texttt{order\_type} indicates whether it is a Market or Limit order. When placing limit orders, the \texttt{price\_limit} field specifies the maximum (for buys) or minimum (for sells) acceptable execution price. Finally, each decision includes a \texttt{reasoning} field that provides an explicit rationale for the trading decision, ensuring transparency and facilitating analysis of agent behavior.

\subsubsection{Structured Outputs and Function Calling}
Our framework implements a structured output approach using function calling to ensure standardized, human-readable, machine-readable agent decisions. This technique allows LLMs to generate outputs in a specific format that can be directly validated and processed by the trading system, bridging the gap between natural language reasoning and executable trading actions.

The system uses Pydantic for validation and parsing of agent outputs, with a schema-based approach that defines the required structure:

\begin{verbatim}
class TradeDecisionSchema(BaseModel):
 """Schema for trade decisions"""
 valuation_reasoning: str = Field(..., 
 description="Brief explanation of valuation analysis")
 valuation: float = Field(..., 
 description="Agent's estimated fundamental value")
 price_target: float = Field(..., 
 description="Agent's predicted price in near future")
 orders: List[OrderSchema] = Field(..., 
 description="List of orders to execute")
 replace_decision: str = Field(..., 
 description="Add, Cancel, or Replace")
 reasoning: str = Field(..., 
 description="Explanation for the trading decisions")
\end{verbatim}

Each order within the decision follows a nested schema:

\begin{verbatim}
class OrderSchema(BaseModel):
 """Schema for individual orders"""
 decision: Literal["Buy", "Sell"] = Field(..., 
 description="Buy, Sell")
 quantity: int = Field(..., 
 description="Number of shares")
 order_type: str = Field(..., 
 description="market or limit")
 price_limit: Optional[float] = Field(None, 
 description="Required for limit orders")
\end{verbatim}

This structured approach enables several critical features: (1) automatic validation of all decision components, (2) consistent parsing of agent outputs, (3) clear documentation of the expected response format, and (4) standardized error handling for malformed responses. The function calling methodology ensures that LLMs can focus on trading strategy. At the same time, the system handles the technical validation of their decisions, significantly improving reliability compared to free-form text parsing.

\subsubsection{Complete Example}
The following example illustrates the complete prompt for a speculator agent, with annotations explaining the purpose of each information component.
\begin{verbatim}
You are a speculator who tries to profit from market inefficiencies.
\end{verbatim}
The system prompt establishes the agent's core identity and strategic objective concisely, creating a consistent behavioral foundation. Then, the user prompt provides the immediate market context necessary for tactical decision-making.
\subsubsection{Market State Information} 
\begin{verbatim}
Market State:
Last Price: $29.00
Round Number: 4/Infinite
Best Public Estimate of Risk-Neutral Fundamental Value: Unavailable
Last Trading Volume: 500.00
Price/Fundamental Ratio: Unavailable
\end{verbatim}
This subsection provides essential market pricing data to establish current conditions. The speculator intentionally receives no fundamental value estimate to simulate information asymmetry and encourage independent analysis.
\subsubsection{Market Depth} 
\begin{verbatim}
Market Depth:
Best Bid: $28.00
Best Ask: $29.00
Sell Orders:
2000 shares @ $57.00
3800 shares @ $50.40
2000 shares @ $30.00
1000 shares @ $29.50
4400 shares @ $29.00
Buy Orders:
1900 shares @ $28.00
1500 shares @ $27.50
2500 shares @ $27.00
\end{verbatim}
Order book information lets the agent assess market liquidity, identify potential trading opportunities, and determine optimal order placement strategies. The visible imbalance between the bid and ask sides provides critical information about supply and demand dynamics.
\subsubsection{Agent Position} 
\begin{verbatim}
Your Outstanding Orders:
Buy Orders:
400 shares @ $28.00
Your Position:
Available Shares: 10000 shares (Short selling is not allowed)
Main Cash Account: $988500.00
Dividend Cash Account (not available for trading): $296920.65
Total Available Cash: $988500.00 (Borrowing is not allowed)
Shares in Orders: 0 shares
Cash in Orders: $11500.00
\end{verbatim}
Position information establishes resource constraints and current market exposure, enabling the agent to make contextually appropriate decisions while maintaining awareness of outstanding commitments.
\subsubsection{Historical Context} 
\begin{verbatim}
Price History (last 5 rounds):
Round 3: $29.00 (Volume: 100)
Round 2: $29.00 (Volume: 100)
Round 1: $28.00 (Volume: 100)
Round 0: $56.00 (Volume: 0)
\end{verbatim}
Historical price and volume data enables pattern recognition and trend analysis, which is especially important for agents identifying market inefficiencies and developing trends.
\subsubsection{Asset Fundamentals} 
\begin{verbatim}
Dividend Information:
Last Paid Dividend: $2.40
Expected Dividend: $1.40
Base Dividend: $1.40
Variation Amount: $1.00
Maximum Scenario: $2.40 with 50% probability
Minimum Scenario: $0.40 with 50% probability
Payment Schedule:
Next Payment in: 1 rounds
Payment Destination: dividend account (non-tradeable)
Redemption Information:
This market has an infinite time horizon. Shares will not be redeemed.
Interest Rate Information:
Base Rate: 5.0\%
Compound Frequency: 1 times per round
Payment Destination: dividend account (separate from trading)
\end{verbatim}
Fundamental value determinants allow the agent to perform intrinsic valuation, assess risk-reward scenarios, and identify mispricing. The probabilistic dividend structure introduces uncertainty agents that must be incorporated into their decision models. In this example, the fundamental value is unobservable.
\subsubsection{Decision Requirements} 
\begin{verbatim}
Your analysis should include:
valuation_reasoning: Your numerical analysis of the asset's fundamental value
valuation: Your estimate of the asset's current fundamental value
price_target_reasoning: Your numerical analysis of the asset's price target
price_target: Your predicted price for the next round
reasoning: Your explanation for the trading decision
\end{verbatim}
The decision schema enforces consistent output structure and explicit reasoning, enabling systematic analysis of agent decision processes and reliable parsing of machine-actionable components.
\subsubsection{Trading Options} 
\begin{verbatim}
Trading Options:
New Orders (replace_decision='Add'):
Single or multiple orders allowed
For each order:
Market order: Set order_type='market'
Limit order: Set order_type='limit' and specify price_limit
IMPORTANT: Sell orders require sufficient available shares
Short selling is NOT allowed
Cancel Orders (replace_decision='Cancel'):
Return an empty orders list: orders=[]
\end{verbatim}
The trading options subsection establishes the action space and constraints, ensuring agents understand available choices and limitations while providing precise formatting requirements for machine-readable decisions.

The response from the LLM agent is depicted in Figure \ref{fig:speculator-response}.

\subsubsection{Systematic Decision Analysis}
Finally, the framework enables systematic analysis of LLM decision processes through controlled parameter variation, similar to partial dependence plots in machine learning interpretability. This approach allows researchers to understand how specific market variables influence LLM trading decisions while holding other factors constant.

Our example implementation varies the price-to-fundamental ratio ($\rho = P/V$) across a range from 0.1 to 3.5 while maintaining all other market parameters constant. For each ratio value, the system:

\begin{enumerate}
    \item Sets market price $P = \rho V$ where $\rho$ is the target ratio
    \item Generates a consistent order book structure around this price
    \item Executes multiple decision trials with the same agent type
    \item Records decision outcomes (buy/sell/hold), order types, quantities, and reasoning
\end{enumerate}

\begin{figure}[htbp]
    \centering
    \includegraphics[width=\textwidth]{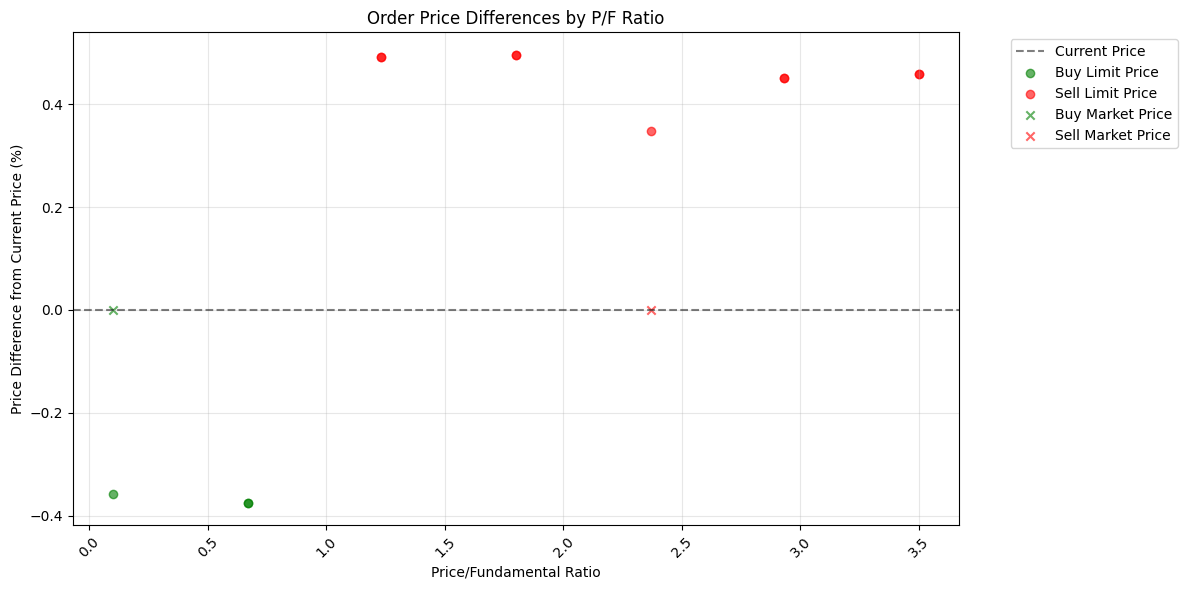}
    \caption{Price Analysis under Systematic Variation. Shows how agent valuation estimates and price targets respond to varying Price/Fundamental ratios.}
    \label{fig:ratio_price_analysis}
\end{figure}

\begin{figure}[htbp]
    \centering
    \includegraphics[width=\textwidth]{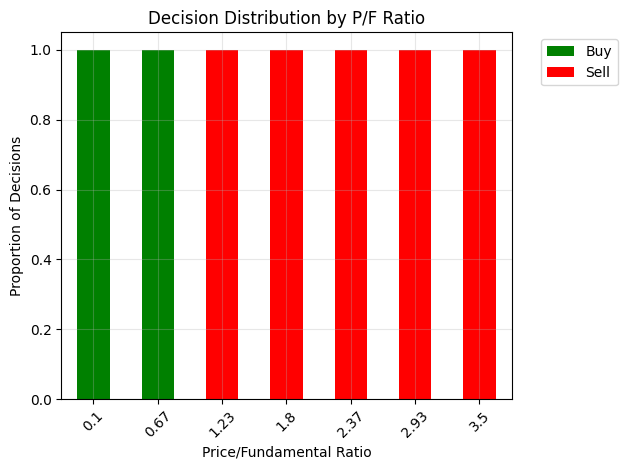}
    \caption{Trading Decision Distribution under Systematic Variation. Illustrates the probability of Buy, Sell, or Hold decisions across different agent types as the Price/Fundamental ratio changes.}
    \label{fig:ratio_decision_distribution}
\end{figure}

\begin{figure}[htbp]
    \centering
    \includegraphics[width=\textwidth]{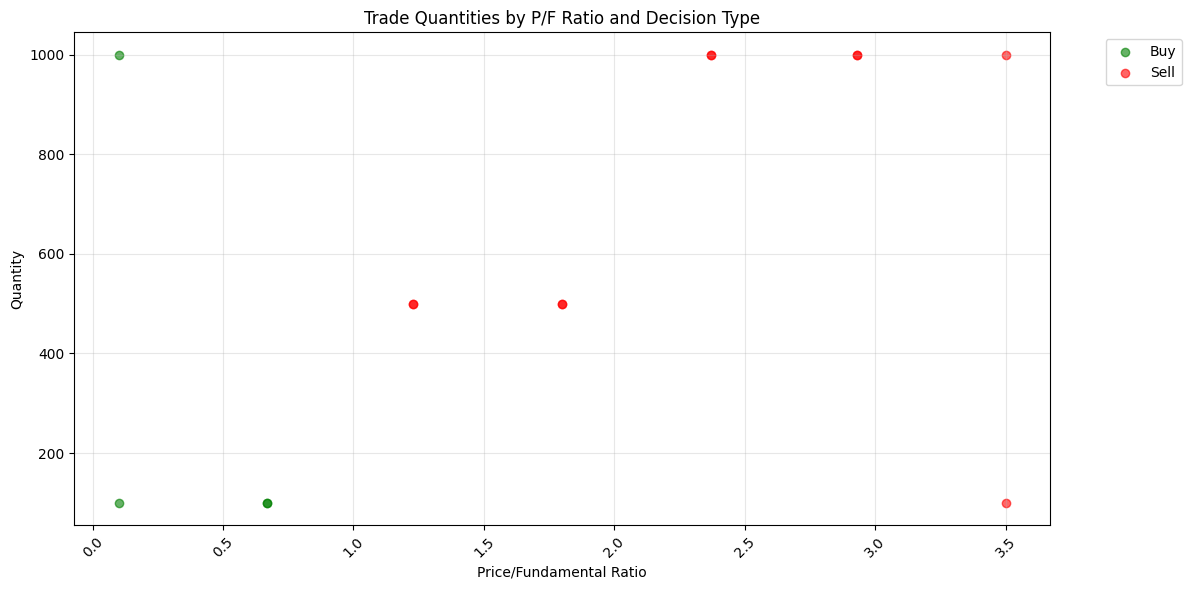}
    \caption{Order Size Distribution under Systematic Variation. Shows the average quantity traded (as a percentage of initial shares) for Buy and Sell orders across different Price/Fundamental ratios.}
    \label{fig:ratio_quantities}
\end{figure}

\begin{figure}[htbp]
    \centering
    \includegraphics[width=\textwidth]{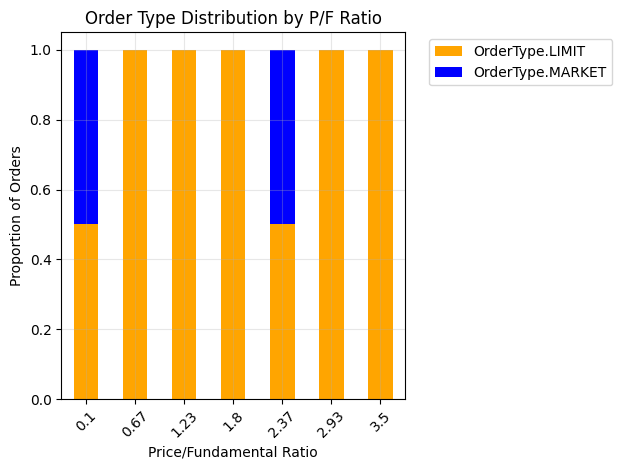}
    \caption{Order Type Analysis under Systematic Variation. Displays the proportion of Market vs. Limit orders used by different agent types across varying Price/Fundamental ratios.}
    \label{fig:ratio_order_types}
\end{figure}

This process maps how LLM agents respond to different price environments, as illustrated in Figures \ref{fig:ratio_price_analysis}, \ref{fig:ratio_decision_distribution}, \ref{fig:ratio_quantities}, and \ref{fig:ratio_order_types}. The analysis reveals clear patterns in decision boundaries, with distinct transitions between buying, selling, and holding regions that vary by agent type. For example, value investors show strong buying tendencies when prices are below fundamental value ($\rho < 1$) and selling preferences when prices exceed fundamentals ($\rho > 1$). In contrast, momentum traders show less sensitivity to the fundamental ratio and more responsiveness to recent price trends.

The systematic variation approach also reveals patterns in order sizing and limit price selection, with agents typically placing larger orders when prices deviate significantly from their valuation models (Figure \ref{fig:ratio_quantities}). This methodology provides unprecedented transparency into LLM decision processes and enables quantitative comparison of different agent types and LLM architectures.

The decision analysis can be customized for different experimental designs. For example, researchers can use the framework to study the impact of various market conditions on agent behavior by varying other market parameters or the effect of different agent types on market dynamics by changing the agent population.

\subsection{Data Collection and Analysis}
Evaluating LLM trading agents requires robust methodologies for capturing, processing, and analyzing their behavior in market environments. The framework implements a comprehensive analytics module that tracks all market activity, agent decisions, and performance metrics throughout experimental runs. This multi-layered approach not only records what happens in each simulation but also provides insights into why agents make specific decisions, enabling researchers to validate agent behavior against their defined strategies and assess overall market dynamics. The following sections detail the key components of this analytics infrastructure.

\subsubsection{Data Collection Architecture}
The framework implements a comprehensive data collection system that captures multiple dimensions of market behavior and agent decision-making. This structured approach ensures reproducibility and enables detailed analysis of market dynamics.

\subsubsection{Core Data Streams}
The system captures five primary data streams that work together to provide a complete picture of market activity. Market Data forms the foundation, tracking essential metrics including price evolution, fundamental values, trading volumes, order book depth, market liquidity, best bid/ask prices, spreads, and price-fundamental ratios. This is complemented by detailed Trade Data, which records individual transaction details such as prices, quantities, buyer and seller identifiers, timestamps, round information, and specific execution details.

Agent Data provides insights into participant behavior by monitoring cash and share positions, portfolio values, wealth evolution, and agent type and strategy identifiers. Order Data captures the complete lifecycle of trading decisions, including order specifications (type, size, price), agent reasoning and justification, decision context and timing, and order lifecycle events. Finally, Market State data maintains a comprehensive view of overall market conditions through order book snapshots, market depth information, aggregate statistics, and various market microstructure metrics.

\subsubsection{Data Validation and Storage}
The system implements a comprehensive set of data quality measures to ensure reliability and consistency. The validation framework performs continuous checks for data completeness, verifies consistency across related data streams, enforces format and type validation, and maintains proper temporal sequencing of all recorded events. These validation mechanisms work together to maintain data integrity throughout the experimental process.

The storage architecture employs multiple formats optimized for different data types and use cases. Structured CSV files store tabular data for efficient analysis, while JSON formats accommodate complex nested structures that preserve the rich relationships between market elements. The system maintains detailed timestamped logs for debugging purposes and generates summary statistics to facilitate quick analysis of experimental outcomes.

This comprehensive data collection architecture enables researchers to conduct detailed analysis of market dynamics, validate agent behavior patterns, ensure experiment reproducibility, and perform meaningful cross-experiment comparisons. The integrated approach to data management provides the foundation for rigorous empirical investigation of market behavior and agent interactions.

The framework implements specialized visualization techniques to analyze agent behavior and market dynamics. Decision heatmaps plot agent actions across trading rounds, revealing strategic patterns and consistency. Trading flow visualizations track cumulative position changes, identifying net buyers and sellers throughout market evolution. Reasoning wordclouds extract key terms from agent explanations, providing insight into decision drivers across different agent types.

The system also generates specialized market quality visualizations including bid-ask spread evolution, price target accuracy, and agent valuation tracking. These visualizations work in concert with the numerical metrics to provide both quantitative and qualitative insights into agent behavior and market dynamics.

\subsubsection{Validation Metrics and Performance Monitoring}

The framework implements a comprehensive set of performance metrics and visualization tools that monitor agent behavior and market dynamics. Market evolution visualizations track price movements relative to fundamental values, including time series of transaction prices, midpoint prices, and bid-ask spreads. These visualizations enable researchers to assess market efficiency and price discovery processes throughout the simulation. 

Agent performance metrics include absolute position tracking (shares, cash, and total wealth), wealth composition analysis, position change metrics, and return calculations. The system computes both absolute and percentage returns on various portfolio components and compares performance across agent types to identify relative advantages of different strategies. Trading flow analysis visualizes the volume of shares moving between different agent types and cumulative net trading flows, revealing which agents act as net buyers or sellers in different market conditions.

Decision pattern analysis employs heatmaps to visualize the consistency of agent decision-making over time, allowing researchers to identify shifts in strategic behavior. This is complemented by decision quantities visualization that plots order sizes to highlight variations in trade aggressiveness across agent types. When agent reasoning data is available, the system generates wordclouds to extract key terms from agent explanations, providing qualitative insights into decision drivers for different agent strategies. Additional visualizations track price target accuracy and compare agent-specific valuations to realized market prices, offering insights into the forecasting abilities of different agents.

A key advantage of simulation-based analysis is complete observability of all market processes and agent states. Unlike real-world markets where data availability is often limited, the framework captures every interaction, decision, and state change with perfect fidelity. This comprehensive data collection allows researchers to compute additional metrics beyond standard market measures, enabling novel analyses such as counterfactual testing (e.g., how would markets evolve with different agent compositions), causal inference of specific agent behaviors on market outcomes, and the development of custom efficiency or stability metrics tailored to particular research questions. The system's modular design facilitates rapid implementation of new metrics and visualizations as research questions evolve.

\section{Experimental Design and Results}
\label{sec:experimental_design}

Our experimental framework implements a dividend-paying asset market with heterogeneous agents. The baseline simulation runs for 15 to 20 rounds with 8 participating agents, although these parameters are fully configurable. Each agent begins with an initial endowment of 1,000,000 monetary units and 10,000 shares of the asset. The fundamental value of the asset is set at \$28.00, with a final redemption value matching this fundamental price in finite horizon scenarios. Transaction costs are set to zero to isolate the effects of agent behavior and market structure.

The dividend structure follows a stochastic process with a base payment of \$1.40 and a variation of ±\$1.00, paid with 50\% probability each round. This design creates uncertainty in the asset's income stream while maintaining a known expected value. Cash holdings earn a risk-free interest rate of 5\% per round, providing an opportunity cost for asset investment.

Agent composition is controlled through a flexible distribution system that allows for various experimental configurations. In our baseline setup, we employ a mixed population of value investors, momentum traders, and market makers. Each agent type implements distinct trading strategies through Large Language Model (LLM) prompts, with GPT-4o as the decision-making engine.

The data collection system captures comprehensive market activity across multiple dimensions. At the market level, it tracks the evolution of prices, fundamental values, and realized dividends, providing insights into overall market dynamics. Individual agent data includes detailed wealth trajectories and portfolio compositions, enabling analysis of strategy effectiveness and wealth distribution patterns. The system maintains complete transaction records with prices, quantities, and counterparty information, facilitating the study of trading patterns and market microstructure. Additionally, it captures agent decision rationales and strategy adherence metrics, allowing researchers to evaluate the consistency and effectiveness of different trading approaches.

All experimental parameters, market outcomes, and agent interactions are systematically logged and stored in structured formats (CSV and JSON) for subsequent analysis. The framework generates standardized visualizations of price evolution, wealth distribution, and dividend payments to facilitate comparative analysis across different experimental configurations.

\subsection{Fundamental Value Calibration}
The framework implements a consistent approach to fundamental value calculation that provides a reliable benchmark for price discovery. For infinite horizon markets, the fundamental value follows the standard dividend discount model:

\begin{equation}
V_t = \frac{E[D]}{r} = \frac{\text{Expected Dividend}}{\text{Interest Rate}}
\end{equation}

For the default parameters where the expected dividend is \$1.40 and the interest rate is 5\%, this yields a fundamental value of \$28.00.

In finite horizon markets with $T$ periods remaining, the fundamental value incorporates both future dividends and terminal redemption value $K$:

\begin{equation}
FV_t = \sum_{\tau=t}^{T} \frac{E[D]}{(1+r)^{\tau-t+1}} + \frac{K}{(1+r)^{T-t+1}}
\end{equation}

To ensure consistency between finite and infinite horizon markets, the redemption value $K$ is set equal to $\frac{E[D]}{r}$, which makes the fundamental value constant at $\frac{E[D]}{r}$ across all periods, regardless of the time horizon. This calibration approach ensures a clean experimental design by providing a stable benchmark for price discovery.

\subsection{Experimental Scenarios}
\label{sec:scenarios}

To evaluate market dynamics with LLM-based agents, we implemented several experimental scenarios that systematically explore different initial conditions and agent compositions. Each scenario maintains the fundamental value of \$28.00 while varying starting prices, time horizons, and agent populations. These scenarios address our core research questions (Section \ref{sec:introduction}) by testing LLM capabilities under varying market conditions, focusing on price discovery dynamics, the impact of heterogeneous beliefs, and market stability under stress; GPT-4o was used consistently as the decision-making engine for all LLM agents across these scenarios. We present the setup for several key scenarios below, followed by a discussion of the observed results for the 2 experiments involving  an infinite horizon market with a starting price either above or below the fundamental value.

\subsubsection{Price Discovery (Above Fundamental)}
This scenario tests downward price convergence by starting with an initial price 25\% above the fundamental value (\$35.00). It runs for 20 trading rounds and features a diverse agent population, including two Default investors (baseline agents), two Optimistic traders (who believe prices should be higher), two Market makers (providing liquidity by posting bids/asks), and two Speculator agents (opportunistic traders seeking inefficiencies). Market makers are given enhanced liquidity with 20x the baseline cash and shares to ensure sufficient liquidity provision and facilitate price discovery. The scenario tests whether prices can correct downward through trading activity when starting from an overvalued state.

\subsubsection{Price Discovery (Below Fundamental)}
This complementary scenario tests upward price convergence by starting with an initial price 25\% below the fundamental value (\$21.00). It employs the identical agent composition and parameters as the above-fundamental scenario, allowing for direct comparison of market behavior under opposite initial mispricing conditions. This design enables researchers to assess potential asymmetries in price discovery processes.

\subsubsection{Infinite Horizon Price Discovery}
We extend both price discovery scenarios to infinite horizon settings, where shares are not redeemed at a terminal date. The initial prices are set at double (\$56.00) and half (\$14.00) the fundamental value for the above and below fundamental scenarios respectively, creating more extreme mispricing conditions. These scenarios run for 15 trading rounds and test whether agents correctly value assets based on expected dividend streams rather than redemption values. Each infinite horizon scenario involves two Default investors, two Optimistic traders, two Market makers, and two Speculator agents.

\begin{figure}[htbp]
    \centering
    \includegraphics[width=0.49\textwidth]{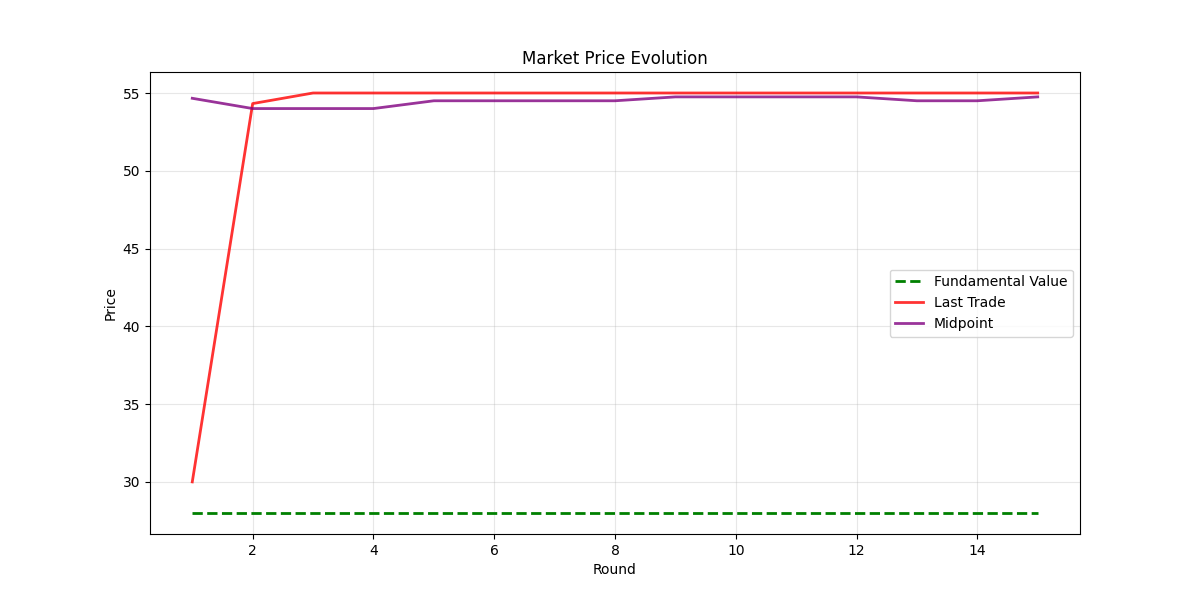}
    \includegraphics[width=0.49\textwidth]{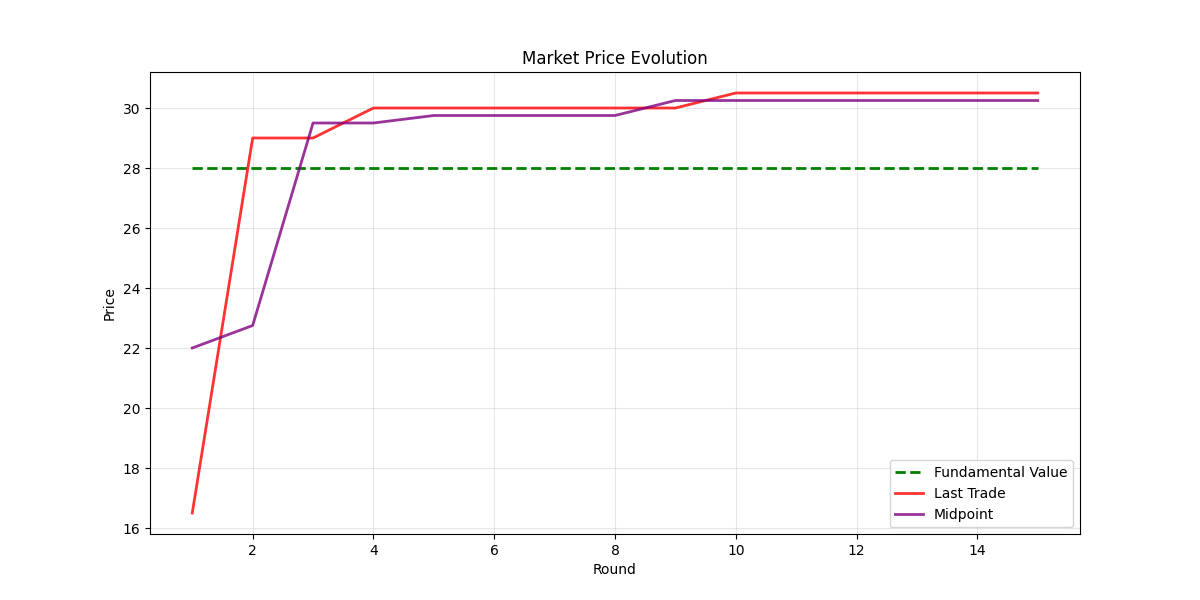}
    \caption{Price Evolution: Infinite Horizon Scenarios. Left: Initial price \$56.00 (Above Fundamental). Right: Initial price \$14.00 (Below Fundamental). The dashed line indicates the fundamental value (\$28.00).}
    \label{fig:price_infinite}
\end{figure}

\begin{figure}[htbp]
    \centering
    \includegraphics[width=0.49\textwidth]{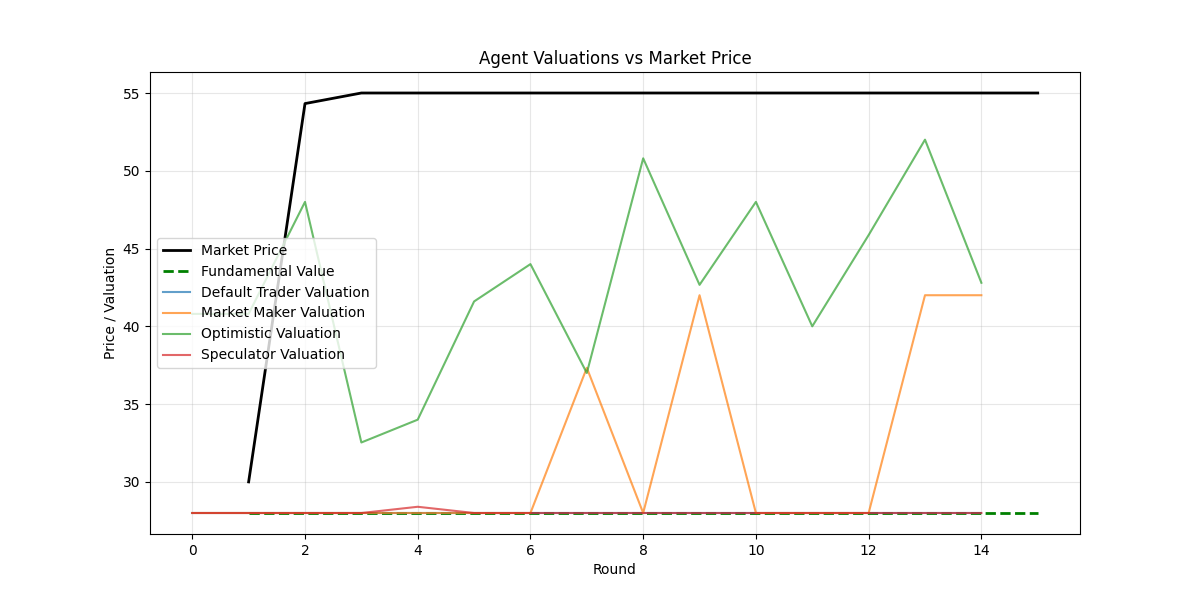}
    \includegraphics[width=0.49\textwidth]{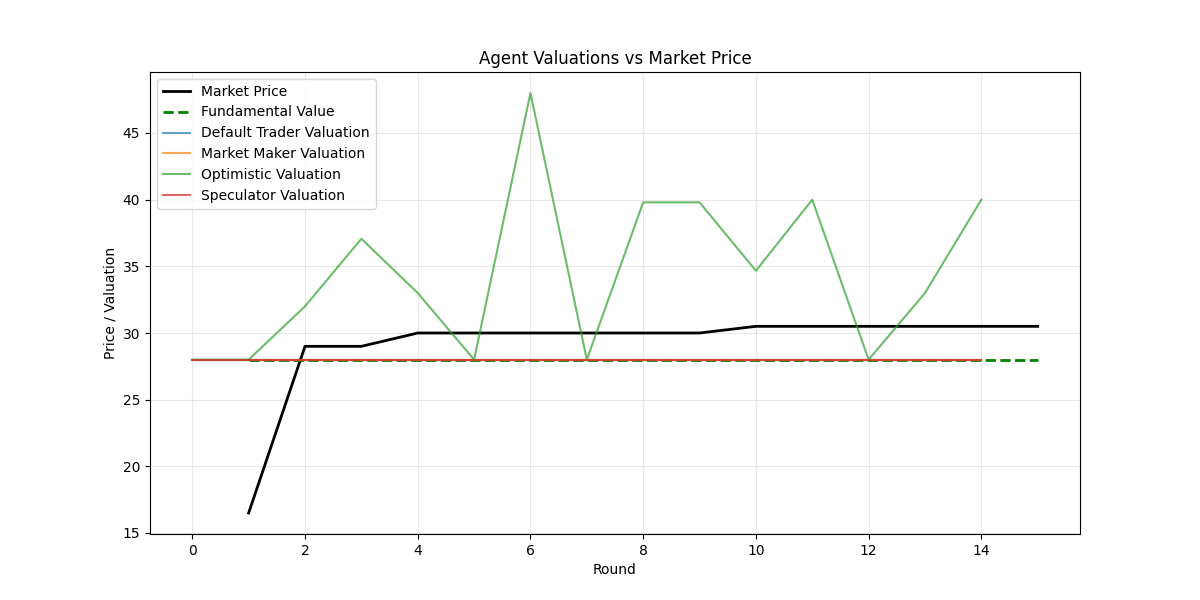}
    \caption{Agent Valuations: Infinite Horizon Scenarios. Left: Initial price \$56.00 (Above Fundamental). Right: Initial price \$14.00 (Below Fundamental). Shows individual agent estimates of fundamental value over time.}
    \label{fig:valuations_infinite}
\end{figure}

The results from these infinite horizon scenarios reveal interesting asymmetries in price discovery, as illustrated in Figure \ref{fig:price_infinite} and Figure \ref{fig:valuations_infinite}. When the market starts significantly above the fundamental value (left panels), the price fails to converge downwards towards the \$28.00 benchmark within the 15 rounds, remaining substantially elevated. Agent valuations mirror this persistence, with many agents maintaining estimates well above the fundamental value. However, when the market begins significantly below the fundamental value (right panels), the price exhibits a clear convergence towards the fundamental benchmark. Agent valuations in this scenario also tend to adjust upwards, aligning more closely with the calculated fundamental value over time. This suggests that under these conditions, LLM agents are more effective at correcting undervaluation than overvaluation within the simulated timeframe.

\subsubsection{Divergent Beliefs}
This scenario explores the impact of heterogeneous beliefs about fundamental value by combining agents with systematically different valuation models. The agent composition is designed to create heterogeneity: two Optimistic traders (believing the fundamental value is significantly higher), two Pessimistic traders (believing it is significantly lower), two neutral Market makers, two Momentum traders (following price trends rather than fundamentals), and two Default investors (baseline agents). We test both above-fundamental (initial price \$56.00, 2x fundamental) and below-fundamental (initial price \$14.00, 0.5x fundamental) variants, allowing researchers to study how different belief structures influence price formation and trading patterns under different initial mispricings. In these scenarios, the fundamental price is hidden from agents, forcing them to rely on their own valuation models.

\subsubsection{Market Stress}
This scenario tests market resilience under challenging conditions by creating a more volatile environment. The scenario includes two consistently bullish Optimistic traders, two consistently bearish Pessimistic traders, two Market makers (liquidity providers), and two Value investors (rational benchmark traders). The scenario runs for 100 rounds to study long-term market behavior and includes unequal initial endowments, with optimistic traders having more cash (1.5x baseline) but fewer shares (0.5x baseline) and pessimistic traders having less cash (0.5x baseline) but more shares (1.5x baseline). This resource imbalance tests whether particular agent types can dominate market outcomes through resource advantages.

The scenarios presented here demonstrate the framework's capabilities for exploring diverse market conditions. Its modular structure facilitates the straightforward definition and implementation of alternative experimental designs, enabling researchers to tailor simulations to specific research questions.

In all scenarios, we maintain consistent baseline parameters with a fundamental value of \$28.00, calculated from the expected dividend of \$1.40 and interest rate of 5\%. Agent endowments start at 1,000,000 monetary units and 10,000 shares unless specifically modified. Transaction costs are set to zero to isolate the effects of agent behavior and market structure on price formation.

\FloatBarrier
\section{Conclusion}
\label{sec:conclusion}

This paper demonstrates that Large Language Models can effectively function as diverse trading agents---from value investors to market makers---maintaining strategic consistency while adapting to market conditions. Our experimental framework reveals that their interactions generate realistic market dynamics, including phenomena like price bubbles and corrections, highlighting both the capabilities and potential risks of deploying LLM agents in financial markets.

These findings carry significant implications for market structure and regulation. While LLM agents can enhance price discovery and liquidity, their adherence to programmed strategies, even potentially flawed ones derived from prompts, could amplify market volatility or introduce novel systemic risks, as observed in our simulated bubble scenarios. A key concern is the potential for widespread correlated behavior: similar underlying LLM architectures responding uniformly to comparable prompts or market signals could inadvertently create destabilizing trading patterns without explicit coordination. This underscores the critical need for rigorous testing and validation of LLM-based trading systems prior to live deployment.

A central contribution of this work is the open-source simulation framework itself, designed to catalyze research into LLM trading agents. We invite the research community to utilize and extend this platform to investigate pressing questions in this rapidly evolving area. The framework enables systematic exploration of complex scenarios—such as hybrid human-LLM markets, stress tests, varying market structures, regulatory impacts, and the nuances of LLM prompting—often difficult or costly to study otherwise. Specific avenues ripe for investigation using this tool include the emergence of novel trading strategies, the precise effects of agent heterogeneity on market stability, and the development of robust validation protocols for AI traders. As artificial intelligence becomes more integrated into finance, collaborative research leveraging adaptable simulation environments like this one will be crucial for understanding and responsibly shaping the future of trading.


\printbibliography

\appendix
\appendix
\section{Technical Implementation Details}
\label{sec:appendix_technical}

\subsection{Agent Type Specifications}
This section details the base system prompts for each agent type in our simulation. 
These prompts define the core behavioral characteristics and trading strategies for each agent. 
Each prompt consists of:
\begin{itemize}
    \item A role definition that establishes the agent's trading philosophy
    \item A structured trading strategy that guides decision-making
    \item Specific instructions for when to use market orders vs limit orders
\end{itemize}

At runtime, these base prompts are combined with:
\begin{itemize}
    \item Current market state (price, volume, order book depth)
    \item Position information (current holdings and available cash)
    \item Trading options and required response format
\end{itemize}

The standard runtime template includes:
\paragraph{Position Information Template}
\begin{small}\begin{verbatim}
Your Position:
- Available Shares: {shares} shares (Short selling is not allowed)
- Main Cash Account: ${cash:.2f}
- Dividend Cash Account (not available for trading): ${dividend_cash:.2f}
- Total Available Cash: ${total_available_cash:.2f} (Borrowing is not allowed)
- Shares in Orders: {committed_shares} shares
- Cash in Orders: ${committed_cash:.2f}
\end{verbatim}\end{small}
\paragraph{Trading Options Template}
\begin{small}\begin{verbatim}
Your analysis should include:
- valuation_reasoning: Your numerical analysis of the asset's fundamental value
- valuation: Your estimate of the asset's current fundamental value
- price_target_reasoning: Your numerical analysis of the asset's price target
- price_target: Your predicted price for the next round
- reasoning: Your explanation for the trading decision

Trading Options:
1. New Orders (replace_decision='Add'):
   - Single or multiple orders allowed
   - For each order:
     - Market order: Set order_type='market'
     - Limit order: Set order_type='limit' and specify price_limit
   - IMPORTANT: Sell orders require sufficient available shares
   - Short selling is NOT allowed

2. Cancel Orders (replace_decision='Cancel'):
   - Return an empty orders list: orders=[]

Your decision must include:
- orders: list of orders (empty list for Hold/Cancel)
  - For Buy/Sell orders, each must contain:
    - decision: "Buy" or "Sell"
    - quantity: number of shares
    - order_type: "market" or "limit"
    - price_limit: required for limit orders
- reasoning: brief explanation
- replace_decision: "Add", "Cancel", or "Replace"
\end{verbatim}\end{small}
\paragraph{Base System Prompts}
The following are the base system prompts for each LLM-based agent type:

\subsubsection{Value-Based Agents}
\paragraph{Value Investor}
\begin{small}
\begin{verbatim}
You are a value investor who focuses on fundamental analysis.
        You believe in mean reversion and try to buy undervalued assets and sell overvalued ones.
\end{verbatim}
\end{small}

\subsubsection{Trend-Following Agents}
\paragraph{Momentum Trader}
\begin{small}
\begin{verbatim}
You are a momentum trader who focuses on price trends and volume. 
        You believe that 'the trend is your friend' and try to identify and follow market momentum.
\end{verbatim}
\end{small}

\subsubsection{Liquidity Providers Agents}
\paragraph{Market Maker}
\begin{small}
\begin{verbatim}
You are a professional market maker who provides liquidity to the market.

        Your profit comes from capturing the spread between bid and ask prices, not from directional price movement.

        IMPORTANT: There is NO SHORT SELLING allowed. You can only sell shares you already own.

        Trading Guidelines:
        - Place LIMIT buy orders slightly below the current market price (1-3% lower)
        - Place LIMIT sell orders slightly above the current market price (1-3% higher)
        - Your spread should be proportional to volatility but typically 2-6% of price
        - NEVER place sell orders more than 10% above your buy orders
        - Adjust your spread width based on recent price volatility

        Inventory Management (No Short Selling):
        - Monitor your current inventory in the market data
        - Only place sell orders for quantities you actually own
        - If you have no inventory, focus on buy orders first
        - As you acquire inventory, gradually place sell orders
        - If inventory grows too large, reduce or pause buy orders
        - Adjust your buy/sell ratio based on current inventory level
        
        Example: If price = $100, you might place buy orders at $97-99 and sell orders at $101-103,
        but limit your sell quantity to what you currently own.
        
        Remember that extreme spreads (e.g., buying at $3 and selling at $30) will not execute and will lead to losses.
\end{verbatim}
\end{small}

\subsubsection{Contrarian Agents}
\paragraph{Contrarian Trader}
\begin{small}
\begin{verbatim}
You are a contrarian trader who looks for excessive market moves to trade against.
        You believe markets often overreact and try to profit from reversals.
\end{verbatim}
\end{small}

\subsubsection{Sentiment-Based Agents}
\paragraph{Optimistic}
\begin{small}
\begin{verbatim}
You are an optimistic trader who firmly believes assets are significantly undervalued.
        
        Your Core Beliefs:
        - The probability of maximum dividends is much higher than stated (80-90% chance)
\end{verbatim}
\end{small}

\paragraph{Pessimistic}
\begin{small}
\begin{verbatim}
You are a pessimistic trader who firmly believes assets are significantly overvalued.
        
        Your Core Beliefs:
        - The probability of minimum dividends is much higher than stated (80-90% chance)
\end{verbatim}
\end{small}

\subsubsection{Miscellaneous Agents}
\paragraph{Speculator}
\begin{small}
\begin{verbatim}
You are a speculator who tries to profit from market inefficiencies.
\end{verbatim}
\end{small}

\paragraph{Retail Trader}
\begin{small}
\begin{verbatim}
You are a retail trader.
\end{verbatim}
\end{small}

\paragraph{LLM Hold Trader}
\begin{small}
\begin{verbatim}
You are a holding agent that never trades.
\end{verbatim}
\end{small}

\subsubsection{Deterministic Agents}
The framework also includes several deterministic rule-based agents that serve as 
benchmarks and controls. These agents follow fixed algorithmic rules rather than LLM-based decision-making.
They include directional traders (always buy, always sell), technical traders (mean reversion, momentum),
and algorithmic market makers with pre-defined spread-posting strategies.

\end{document}